\pdfoutput=1
\documentclass[12pt,a4paper]{article}
\usepackage{ifthen} 
\newboolean{pdflatex}
\setboolean{pdflatex}{true} 

\newboolean{articletitles}
\setboolean{articletitles}{true} 

\newboolean{uprightparticles}
\setboolean{uprightparticles}{false}

\newboolean{inbibliography}
\setboolean{inbibliography}{false}

\usepackage{microtype}
\usepackage{lineno}  % for line numbering during review
\usepackage{xspace} % To avoid problems with missing or double spaces after
\usepackage{graphicx}  % to include figures (can also use other packages)
\usepackage{color}
\usepackage{colortbl}
\usepackage[small]{caption}
\graphicspath{{./figs/}} % Make Latex search fig subdir for figures

\usepackage{amsmath} % Adds a large collection of math symbols
\usepackage{amssymb}
\usepackage{amsfonts}
\usepackage{upgreek} % Adds in support for greek letters in roman typeset

\newcommand*\patchAmsMathEnvironmentForLineno[1]{%
\expandafter\let\csname old#1\expandafter\endcsname\csname #1\endcsname
\expandafter\let\csname oldend#1\expandafter\endcsname\csname
end#1\endcsname
 \renewenvironment{#1}%
   {\linenomath\csname old#1\endcsname}%
   {\csname oldend#1\endcsname\endlinenomath}%
}
\newcommand*\patchBothAmsMathEnvironmentsForLineno[1]{%
  \patchAmsMathEnvironmentForLineno{#1}%
  \patchAmsMathEnvironmentForLineno{#1*}%
}
\AtBeginDocument{%
\patchBothAmsMathEnvironmentsForLineno{equation}%
\patchBothAmsMathEnvironmentsForLineno{align}%
\patchBothAmsMathEnvironmentsForLineno{flalign}%
\patchBothAmsMathEnvironmentsForLineno{alignat}%
\patchBothAmsMathEnvironmentsForLineno{gather}%
\patchBothAmsMathEnvironmentsForLineno{multline}%
}

\usepackage{hyperref}    % Hyperlinks in references
\usepackage[all]{hypcap} % Internal hyperlinks to floats.

\def\lhcb {\mbox{LHCb}\xspace}

\def\babar  {\mbox{BaBar}\xspace}
\def\belle  {\mbox{Belle}\xspace}

\def\argus  {\mbox{ARGUS}\xspace}
\ifthenelse{\boolean{uprightparticles}}%
{

 \def\Pmu         {\ensuremath{\upmu}\xspace}

 \def\Ppi         {\ensuremath{\uppi}\xspace}

 \def\Ppsi        {\ensuremath{\uppsi}\xspace}

 \def\PDelta      {\ensuremath{\Delta}\xspace}                 
 \def\PXi      {\ensuremath{\Xi}\xspace}                 
 \def\PLambda      {\ensuremath{\Lambda}\xspace}                 
 \def\PSigma      {\ensuremath{\Sigma}\xspace}                 
 \def\POmega      {\ensuremath{\Omega}\xspace}                 
 \def\PUpsilon      {\ensuremath{\Upsilon}\xspace}                 
 
 \def\PB      {\ensuremath{\mathrm{B}}\xspace}                 
                  
 \def\PD      {\ensuremath{\mathrm{D}}\xspace}

 \def\PJ      {\ensuremath{\mathrm{J}}\xspace}                 
 \def\PK      {\ensuremath{\mathrm{K}}\xspace}

 \def\Pb      {\ensuremath{\mathrm{b}}\xspace}                 
 \def\Pc      {\ensuremath{\mathrm{c}}\xspace}

 \def\Pi      {\ensuremath{\mathrm{i}}\xspace}

}
{

 \def\Pmu         {\ensuremath{\mu}\xspace}

 \def\Ppi         {\ensuremath{\pi}\xspace}

 \def\Ppsi        {\ensuremath{\psi}\xspace}                 
                  
 \mathchardef\PDelta="7101
 \mathchardef\PXi="7104
 \mathchardef\PLambda="7103
 \mathchardef\PSigma="7106
 \mathchardef\POmega="710A
 \mathchardef\PUpsilon="7107
                  
 \def\PB      {\ensuremath{B}\xspace}                 
                  
 \def\PD      {\ensuremath{D}\xspace}

 \def\PJ      {\ensuremath{J}\xspace}                 
 \def\PK      {\ensuremath{K}\xspace}

 \def\Pb      {\ensuremath{b}\xspace}                 
 \def\Pc      {\ensuremath{c}\xspace}

 \def\Pi      {\ensuremath{i}\xspace}

}

\def\mumu       {\ensuremath{\Pmu^+\Pmu^-}\xspace}

\def\cquark    {\ensuremath{\Pc}\xspace}

\def\bquark    {\ensuremath{\Pb}\xspace}

\def\pion  {\ensuremath{\Ppi}\xspace}

\def\pip   {\ensuremath{\pion^+}\xspace}
\def\pim   {\ensuremath{\pion^-}\xspace}

\def\kaon  {\ensuremath{\PK}\xspace}
  \def\Kbar  {\kern 0.2em\overline{\kern -0.2em \PK}{}\xspace}

\def\Kp    {\ensuremath{\kaon^+}\xspace}
\def\Km    {\ensuremath{\kaon^-}\xspace}

\def\Kstarz  {\ensuremath{\kaon^{*0}}\xspace}

  \def\Dbar    {\kern 0.2em\overline{\kern -0.2em \PD}{}\xspace}

\def\Dzb     {\ensuremath{\Dbar^0}\xspace}

\def\B       {\ensuremath{\PB}\xspace}
\def\Bbar    {\ensuremath{\kern 0.18em\overline{\kern -0.18em \PB}{}}\xspace}

\def\Bz      {\ensuremath{\B^0}\xspace}

\def\Bu      {\ensuremath{\B^+}\xspace}
\def\Bub     {\ensuremath{\B^-}\xspace}
\def\Bp      {\ensuremath{\Bu}\xspace}
\def\Bm      {\ensuremath{\Bub}\xspace}

\def\jpsi     {\ensuremath{{\PJ\mskip -3mu/\mskip -2mu\Ppsi\mskip 2mu}}\xspace}
\def\psitwos  {\ensuremath{\Ppsi{(2S)}}\xspace}

  \def\Y#1S{\ensuremath{\PUpsilon{(#1S)}}\xspace}% no space before {...}!

\def\Lbar {\ensuremath{\kern 0.1em\overline{\kern -0.1em\PLambda}}\xspace}

\def\to                 {\ensuremath{\rightarrow}\xspace}

\def\qsq       {\ensuremath{q^2}\xspace}

\def\CP                {\ensuremath{C\!P}\xspace}

\newcommand{\ACP}{\ensuremath{{\cal A}_{\CP}}\xspace}

\def\AT#1     {\ensuremath{A_{\mathrm{T}}^{#1}}\xspace}           % 2

\def\C#1      {\ensuremath{\mathcal{C}_{#1}}\xspace}                       % 9
\def\Cp#1     {\ensuremath{\mathcal{C}_{#1}^{'}}\xspace}                    % 7
\def\Ceff#1   {\ensuremath{\mathcal{C}_{#1}^{\mathrm{(eff)}}}\xspace}        % 9  
\def\Cpeff#1  {\ensuremath{\mathcal{C}_{#1}^{'\mathrm{(eff)}}}\xspace}       % 7
\def\Ope#1    {\ensuremath{\mathcal{O}_{#1}}\xspace}                       % 2
\def\Opep#1   {\ensuremath{\mathcal{O}_{#1}^{'}}\xspace}                    % 7

\newcommand{\tev}{\ifthenelse{\boolean{inbibliography}}{\ensuremath{~T\kern -0.05em eV}\xspace}{\ensuremath{\mathrm{\,Te\kern -0.1em V}}\xspace}}
\newcommand{\gev}{\ensuremath{\mathrm{\,Ge\kern -0.1em V}}\xspace}
\newcommand{\mev}{\ensuremath{\mathrm{\,Me\kern -0.1em V}}\xspace}
\newcommand{\kev}{\ensuremath{\mathrm{\,ke\kern -0.1em V}}\xspace}
\newcommand{\ev}{\ensuremath{\mathrm{\,e\kern -0.1em V}}\xspace}
\newcommand{\gevc}{\ensuremath{{\mathrm{\,Ge\kern -0.1em V\!/}c}}\xspace}
\newcommand{\mevc}{\ensuremath{{\mathrm{\,Me\kern -0.1em V\!/}c}}\xspace}
\newcommand{\gevcc}{\ensuremath{{\mathrm{\,Ge\kern -0.1em V\!/}c^2}}\xspace}
\newcommand{\gevgevcccc}{\ensuremath{{\mathrm{\,Ge\kern -0.1em V^2\!/}c^4}}\xspace}
\newcommand{\mevcc}{\ensuremath{{\mathrm{\,Me\kern -0.1em V\!/}c^2}}\xspace}

\def\mum  {\ensuremath{\,\upmu\rm m}\xspace}

\def\invfb   {\ensuremath{\mbox{\,fb}^{-1}}\xspace}

\newcommand{\chisq}{\ensuremath{\chi^2}\xspace}

\newcommand{\chisqip}{\ensuremath{\chi^2_{\rm IP}}\xspace}

\def\gsim{{~\raise.15em\hbox{$>$}\kern-.85em
          \lower.35em\hbox{$\sim$}~}\xspace}
\def\lsim{{~\raise.15em\hbox{$<$}\kern-.85em
          \lower.35em\hbox{$\sim$}~}\xspace}

\def\pt         {\mbox{$p_{\rm T}$}\xspace}

\def\evtgen     {\mbox{\textsc{EvtGen}}\xspace}

\def\gauss      {\mbox{\textsc{Gauss}}\xspace}
\def\geant      {\mbox{\textsc{Geant4}}\xspace}

\def\photos     {\mbox{\textsc{Photos}}\xspace}

\def\pythia     {\mbox{\textsc{Pythia}}\xspace}

\def\tell1  {TELL1\xspace}
\def\ukl1   {UKL1\xspace}

\def\BToKmumu   {\ensuremath{{\Bu \rightarrow \Kp \mumu}}\xspace}
\def\BToJpsiK   {\ensuremath{{\Bu \rightarrow \jpsi \Kp}}\xspace}
\def\BToKmumubar   {\ensuremath{\Bub \rightarrow \Km \mumu}\xspace}

\def\BToKpipi   {\ensuremath{{\Bu \rightarrow \Kp \pip\pim}}\xspace}
\def\BTopimumu   {\ensuremath{{\Bu \rightarrow \pip \mumu}}\xspace}

\newcommand{\BdToKstarmumu}{\ensuremath{\B^0 \rightarrow K^{*0} \mu^+ \mu^-}\xspace}

\newcommand{\AD}{\ensuremath{\mathcal{A}_{\rm D}}\xspace}
\newcommand{\AP}{\ensuremath{\mathcal{A}_{\rm P}}\xspace}
\newcommand{\ARAW}{\ensuremath{\mathcal{A}_{\rm RAW}}\xspace}

\newcommand{\NCKmumu}{\ensuremath{K \mu \mu}\xspace}

\usepackage{cite} % Allows for ranges in citations
\usepackage{mciteplus}

\newcommand\Tstrut{\rule{0pt}{2.6ex}}

\usepackage{longtable}

\usepackage{dcolumn}

\begin{document}

\renewcommand{\thefootnote}{\fnsymbol{footnote}}
\setcounter{footnote}{1}

\begin{titlepage}
\pagenumbering{roman}

\vspace*{-1.5cm}
\centerline{\large EUROPEAN ORGANIZATION FOR NUCLEAR RESEARCH (CERN)}
\vspace*{1.5cm}
\hspace*{-0.5cm}
\begin{tabular*}{\linewidth}{lc@{\extracolsep{\fill}}r}
\ifthenelse{\boolean{pdflatex}}
\vspace*{-2.7cm}\mbox{\!\!\!\includegraphics[width=.14\textwidth]{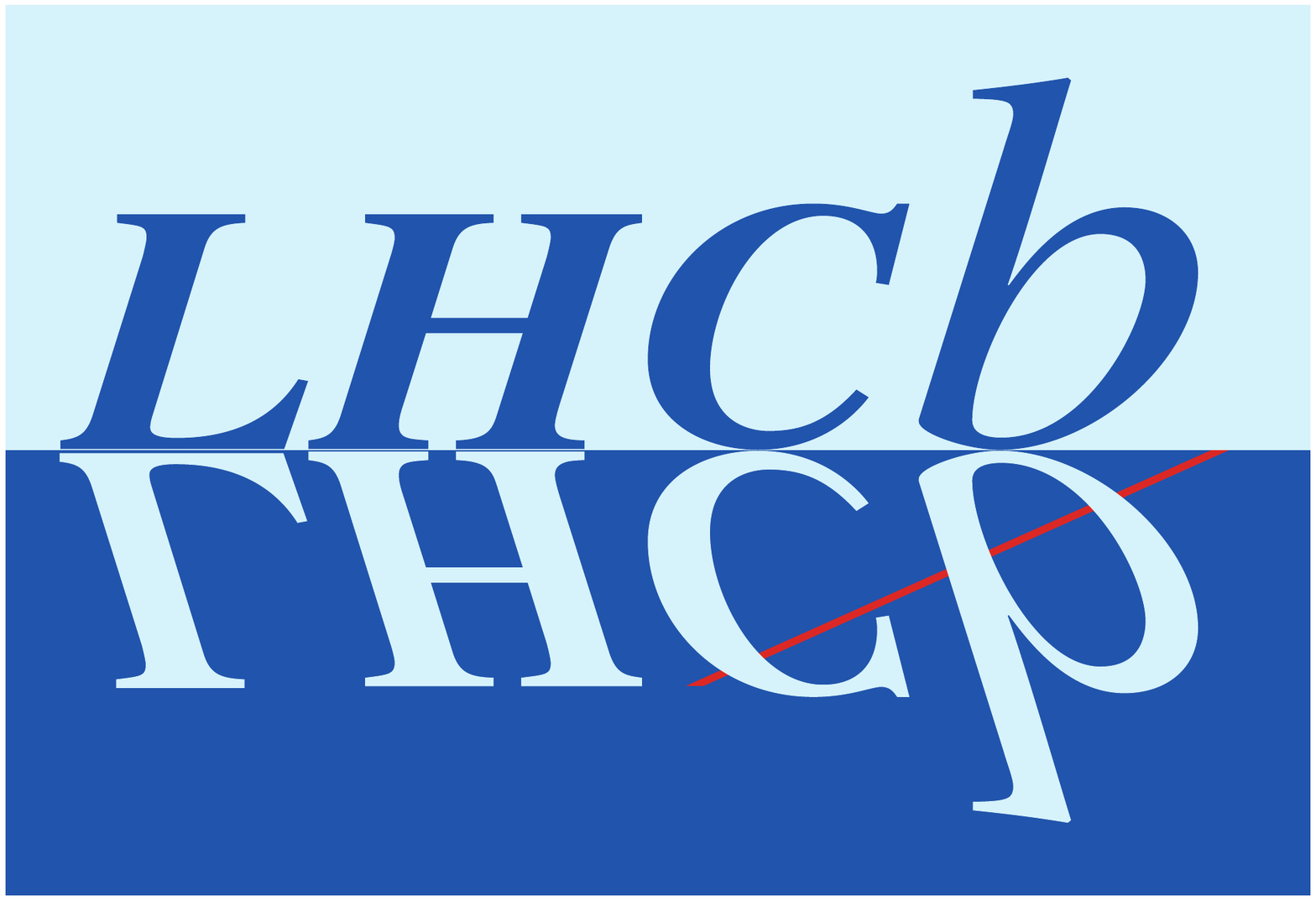}} & &%
%{\vspace*{-1.2cm}\mbox{\!\!\!\includegraphics[width=.12\textwidth]{figs/lhcb-logo.eps}} & &}%
\\
 & & CERN-PH-EP-2013-145 \\  % ID 
 & & LHCb-PAPER-2013-043 \\  % ID 
 & & August 6, 2013 \\
 & & \\
\end{tabular*}

\vspace*{3.0cm}

{\bf\boldmath\huge
\begin{center}
  Measurement of the \CP asymmetry in \BToKmumu decays
\end{center}
}

\vspace*{2.0cm}

\begin{center}
The LHCb collaboration\footnote{Authors are listed on the following pages.}
\end{center}

\vspace{\fill}

\begin{abstract}
  \noindent
A measurement of the \CP asymmetry in \BToKmumu decays is presented using $pp$ collision data, corresponding to an integrated luminosity of 1.0\invfb, recorded by the \lhcb experiment during 2011 at a centre-of-mass energy of 7\tev. The measurement is performed in seven bins of \mumu invariant mass squared in the range ${0.05<\qsq<22.00\gevgevcccc}$, excluding the \jpsi and \psitwos resonance regions. Production and detection asymmetries are corrected for using the \BToJpsiK decay as a control mode. Averaged over all the bins, the \CP asymmetry is found to be ${\ACP = 0.000\pm 0.033\mbox{ (stat.)} \pm0.005 \mbox{ (syst.)} \pm 0.007\mbox{ }(\jpsi\Kp)}$, where the third uncertainty is due to the \CP asymmetry of the control mode. This is consistent with the Standard Model prediction.
\end{abstract}

\vspace*{2.0cm}

\begin{center}
  Published in Phys.~Rev.~Lett.
\end{center}

\vspace{\fill}

{\footnotesize 
\centerline{\copyright~CERN on behalf of the \lhcb collaboration, license \href{http://creativecommons.org/licenses/by/3.0/}{CC-BY-3.0}.}}
\vspace*{2mm}

\end{titlepage}

\newpage
\setcounter{page}{2}
\mbox{~}
\newpage

%AUTHORLIST%
%%%%%%%%%%%%%%%%%%%%%%%%%%%%%%%%%%%%%%%%%%
\centerline{\large\bf LHCb collaboration}
\begin{flushleft}
\small
R.~Aaij$^{40}$, 
B.~Adeva$^{36}$, 
M.~Adinolfi$^{45}$, 
C.~Adrover$^{6}$, 
A.~Affolder$^{51}$, 
Z.~Ajaltouni$^{5}$, 
J.~Albrecht$^{9}$, 
F.~Alessio$^{37}$, 
M.~Alexander$^{50}$, 
S.~Ali$^{40}$
% ...

{\footnotesize \it
% ...
$ ^{5}$Clermont Universit\'{e}, Universit\'{e} Blaise Pascal, CNRS/IN2P3, LPC, Clermont-Ferrand, France\\
$ ^{6}$CPPM, Aix-Marseille Universit\'{e}, CNRS/IN2P3, Marseille, France\\
% ...
$ ^{9}$Fakult\"{a}t Physik, Technische Universit\"{a}t Dortmund, Dortmund, Germany\\
% ...
$ ^{36}$Universidad de Santiago de Compostela, Santiago de Compostela, Spain\\
$ ^{37}$European Organization for Nuclear Research (CERN), Geneva, Switzerland\\
% ...
$ ^{40}$Nikhef National Institute for Subatomic Physics, Amsterdam, The Netherlands\\
% ...
$ ^{45}$H.H. Wills Physics Laboratory, University of Bristol, Bristol, United Kingdom\\
% ...
$ ^{50}$School of Physics and Astronomy, University of Glasgow, Glasgow, United Kingdom\\
$ ^{51}$Oliver Lodge Laboratory, University of Liverpool, Liverpool, United Kingdom\\
% ...
}
\end{flushleft}
%%%%%%%%%%%%%%%%%%%%%%%%%%%%%%%%%%%%%%%%%%
%AUTHORLIST

\cleardoublepage

\renewcommand{\thefootnote}{\arabic{footnote}}
\setcounter{footnote}{0}

\pagestyle{plain} 
\setcounter{page}{1}
\pagenumbering{arabic}

\noindent The rare decay \BToKmumu is a flavour-changing neutral current process mediated by electroweak loop (penguin) and box diagrams. The absence of tree-level diagrams for the decay results in a small value of the Standard Model (SM) prediction for the branching fraction, which is supported by a measurement of ${(4.36 \pm 0.23) \times 10^{-7}}$~\cite{LHCb-PAPER-2012-024}. Physics processes beyond the SM that may enter via the loop and box diagrams could have large effects on observables of the decay. Examples include the decay rate, the \mumu forward-backward asymmetry~\cite{LHCb-PAPER-2012-024,Wei:2009zv,Aaltonen:2011ja}, and the \CP asymmetry~\cite{Wei:2009zv,Babar:2012vwa}, as functions of the \mumu invariant mass squared (\qsq).

The \CP asymmetry is defined as
\begin{equation}
 \ACP = \frac{\Gamma(\BToKmumubar)-\Gamma(\BToKmumu)}{\Gamma(\BToKmumubar)+\Gamma(\BToKmumu)},
\label{eq:ACP}
\end{equation} 
where $\Gamma$ is the decay rate of the mode. This asymmetry is predicted to be of order $10^{-4}$ in the SM~\cite{Bobeth:2011nj}, but can be significantly enhanced in models beyond the SM~\cite{Altmannshofer:2008dz}. Current measurements including the dielectron mode, $\ACP({B \rightarrow \Kp \ell^+ \ell^-})$, from \babar and \belle give ${-0.03 \pm 0.14}$ and ${0.04 \pm 0.10}$, respectively~\cite{Babar:2012vwa,Wei:2009zv}, and are consistent with the SM. The \CP asymmetry has already been measured at \lhcb in \BdToKstarmumu decays~\cite{LHCb:2012kz}, ${\ACP = -0.072 \pm 0.040}$. Assuming that contributions beyond the SM are independent of the flavour of the spectator quark, \ACP should be similar for both \BToKmumu and \BdToKstarmumu decays.

In this Letter, a measurement of \ACP in \BToKmumu decays is presented using $pp$ collision data, corresponding to an integrated luminosity of 1.0\invfb, recorded at a centre-of-mass energy of 7\tev at \lhcb in 2011. The inclusion of charge conjugate modes is implied throughout unless explicitly stated.

The \lhcb detector~\cite{Alves:2008zz} is a single-arm forward
spectrometer covering the \mbox{pseudorapidity} range $2<\eta <5$,
designed for the study of particles containing \bquark or \cquark
quarks. The detector includes a high-precision tracking system
consisting of a silicon-strip vertex detector surrounding the $pp$
interaction region, a large-area silicon-strip detector located
upstream of a dipole magnet with a bending power of about
$4{\rm\,Tm}$, and three stations of silicon-strip detectors and straw
drift tubes placed downstream.
The combined tracking system provides a momentum measurement with
relative uncertainty that varies from 0.4\% at 5\gevc to 0.6\% at 100\gevc,
and impact parameter (IP) resolution of 20\mum for
tracks with high transverse momentum (\pt). Charged hadrons are identified
using two ring-imaging Cherenkov detectors~\cite{LHCb-DP-2012-003}. Muons are identified by a
system composed of alternating layers of iron and multiwire
proportional chambers~\cite{LHCb-DP-2012-002}.

Samples of simulated events are used to determine the efficiency of selecting \BToKmumu signal events and to study certain backgrounds. In the simulation, $pp$ collisions are generated using
\pythia~6.4~\cite{Sjostrand:2006za} with a specific \lhcb
configuration~\cite{LHCb-PROC-2010-056}.  Decays of hadronic particles
are described by \evtgen~\cite{Lange:2001uf}, in which final-state
radiation is generated using \photos~\cite{Golonka:2005pn}. The
interaction of the generated particles with the detector and its
response are implemented using the \geant
toolkit~\cite{Allison:2006ve, *Agostinelli:2002hh} as described in
Ref.~\cite{LHCb-PROC-2011-006}.
The simulated samples are corrected to reproduce the data distributions of the \Bp meson \pt and vertex \chisq, the track \chisq of the kaon, as well as the detector IP resolution, particle identification and momentum resolution.

Candidate events are first required to pass a hardware trigger,
which selects muons with $\pt>1.48\gevc$~\cite{LHCb-DP-2012-004}. In
the subsequent software trigger, at least
one of the final-state particles is required to have 
$\pt>1.0\gevc$ and IP $>100\mum$ with respect to all primary $pp$ interaction vertices~(PVs) in the
event. Finally, the tracks of two or more of the final-state
particles are required to form a vertex that is displaced from the PVs.

An initial selection is applied to the \BToKmumu candidates to enhance signal decays and suppress combinatorial background. Candidate \Bp mesons must satisfy requirements on their direction and flight distance, to ensure consistency with originating from the PV. The decay products must pass criteria regarding the \chisqip, where \chisqip is defined as the difference in \chisq of a given PV reconstructed with and without the considered particle. There is also a requirement on the vertex \chisq of the \mumu pair. All the tracks are required to have \pt $>250\mevc$.

Additional background rejection is achieved by using a boosted decision tree (BDT)~\cite{Breiman} that implements the AdaBoost algorithm~\cite{AdaBoost}. The BDT uses the \pt and \chisqip of the muons and the \Bp meson candidate, as well as the decay time, vertex \chisq, and flight direction of the \Bp candidate and the \chisqip of the kaon. Data, corresponding to an integrated luminosity of 0.1\invfb, are used to optimise this selection, leaving 0.9\invfb for the determination of \ACP.

Following the multivariate selection, candidate events pass several requirements to remove specific sources of background. Particle identification (PID) criteria are applied to kaon candidates to reduce the number of pions incorrectly identified as kaons. Candidates with \mumu invariant mass in the ranges $2.95 < m_{\mu\mu} < 3.18\gevcc$ and $3.59 < m_{\mu\mu} < 3.77\gevcc$ are removed to reject backgrounds from tree level ${\Bu \rightarrow \jpsi (\rightarrow \mumu) \Kp}$ and ${\Bu \rightarrow \psitwos(\rightarrow \mumu) \Kp}$ decays. Those in the first range are selected as \BToJpsiK decays, which are used as a control sample. If $m_{\NCKmumu} < 5.22\gevcc$, the vetoes are extended downwards by 0.25 and 0.19\gevcc, respectively, to remove the radiative tails of the resonant decays. If $5.35 < m_{\NCKmumu} < 5.50\gevcc$ the vetoes are extended upwards by 0.05\gevcc to remove misreconstructed resonant candidates that appear at large $m_{\mu\mu}$ and $m_{\NCKmumu}$. Further vetoes are applied to remove \BToJpsiK events in which the kaon and a muon have been swapped, and contributions from decays involving charm mesons such as ${\Bp \rightarrow \Dzb (\rightarrow \Kp\pim)\pip}$ where both pions are misidentified as muons. After these selection requirements have been applied, there are two sources of background that are difficult to distinguish from the signal. These are \BToKpipi and \BTopimumu decays, which both contribute at the level of 1\% of the signal yield. These peaking backgrounds are accounted for during the analysis.

In order to perform a measurement of \ACP, the production and detection asymmetries associated with the measurement must be considered. The raw measured asymmetry is, to first order,

\begin{equation}
 \ARAW(\BToKmumu) = \ACP(\BToKmumu) + \AP + \AD,
\end{equation}
where the production and detection asymmetries are defined as
\begin{eqnarray}
\AP &\equiv& [R(\Bm) - R(\Bp)]/[R(\Bm) + R(\Bp)], \\
\AD &\equiv& [\epsilon(\Km) - \epsilon(\Kp)]/[\epsilon(\Km) + \epsilon(\Kp)],
\end{eqnarray}
where $R$ and $\epsilon$ represent the $B$ meson production rate and kaon detection efficiency, respectively. The detection asymmetry has two components: one due to the different interactions of positive and negative kaons with the detector material, and a left-right asymmetry due to particles of different charges being deflected to opposite sides of the detector by the magnet. The component of the detection asymmetry from muon reconstruction is small and neglected. Since the \lhcb experiment reverses the magnetic field, about half of the data used in the analysis is taken with each polarity. Therefore, an average of the measurements with the two polarities is used to suppress significantly the second effect. To account for both the detection and production asymmetries, the decay \BToJpsiK is used, which has the same final-state particles as \BToKmumu and very similar kinematic properties. The \CP asymmetry in \BToJpsiK decays has been measured as $(1 \pm 7) \times 10^{-3}$~\cite{PDG2012,Abazov:2008gs}. Neglecting the difference in the final-state kinematic properties of the kaon, the production and detection asymmetries are the same for both modes, and the value of the \CP asymmetry can be obtained via

{\footnotesize
\begin{equation}
 \ACP(\BToKmumu) = \ARAW(\BToKmumu) - \ARAW(\BToJpsiK) + \ACP(\BToJpsiK).
\label{eq:ACP2}
\end{equation}
}
Differences in the kinematic properties are accounted for by a systematic uncertainty.

In the data set, approximately 1330 \BToKmumu and 218,000 \BToJpsiK signal decays are reconstructed. To measure any variation in \ACP as a function of \qsq, which improves the sensitivity of the measurement to physics beyond the SM, the \BToKmumu dataset is divided into the seven \qsq bins used in Ref.~\cite{LHCb-PAPER-2012-024}. The measurement is also made in a bin of $1<\qsq<6\gevgevcccc$, which is of particular theoretical interest. To determine the number of \Bp decays in each bin, a simultaneous unbinned maximum likelihood fit is performed to the invariant mass distributions of the \BToKmumu and \BToJpsiK candidates in the range ${5.10<m_{\NCKmumu}<5.60\gevcc}$. The signal shape is parameterised by a Cruijff function~\cite{delAmoSanchez:2010ae}, and the combinatorial background is described by an exponential function. All parameters of the signal and combinatorial background are allowed to vary freely in the fit. Additionally, there is background from partially-reconstructed decays such as ${\Bz \rightarrow \Kstarz (\rightarrow \Kp\pim) \mumu}$ or ${\Bz \rightarrow \jpsi \Kstarz(\rightarrow \Kp\pim)}$ where the pion is undetected. For the \BToKmumu distribution, these decays are fitted by an \argus function~\cite{Albrecht:1989ga} convolved with a Gaussian function to account for detector resolution. For the \BToJpsiK decays the partially-reconstructed background is modelled by another Cruijff function. The shapes of the peaking backgrounds, due to \BToKpipi and \BTopimumu decays, are taken from fits to simulated events.

\begin{figure}[tb]
\centering
\includegraphics[width=\textwidth]{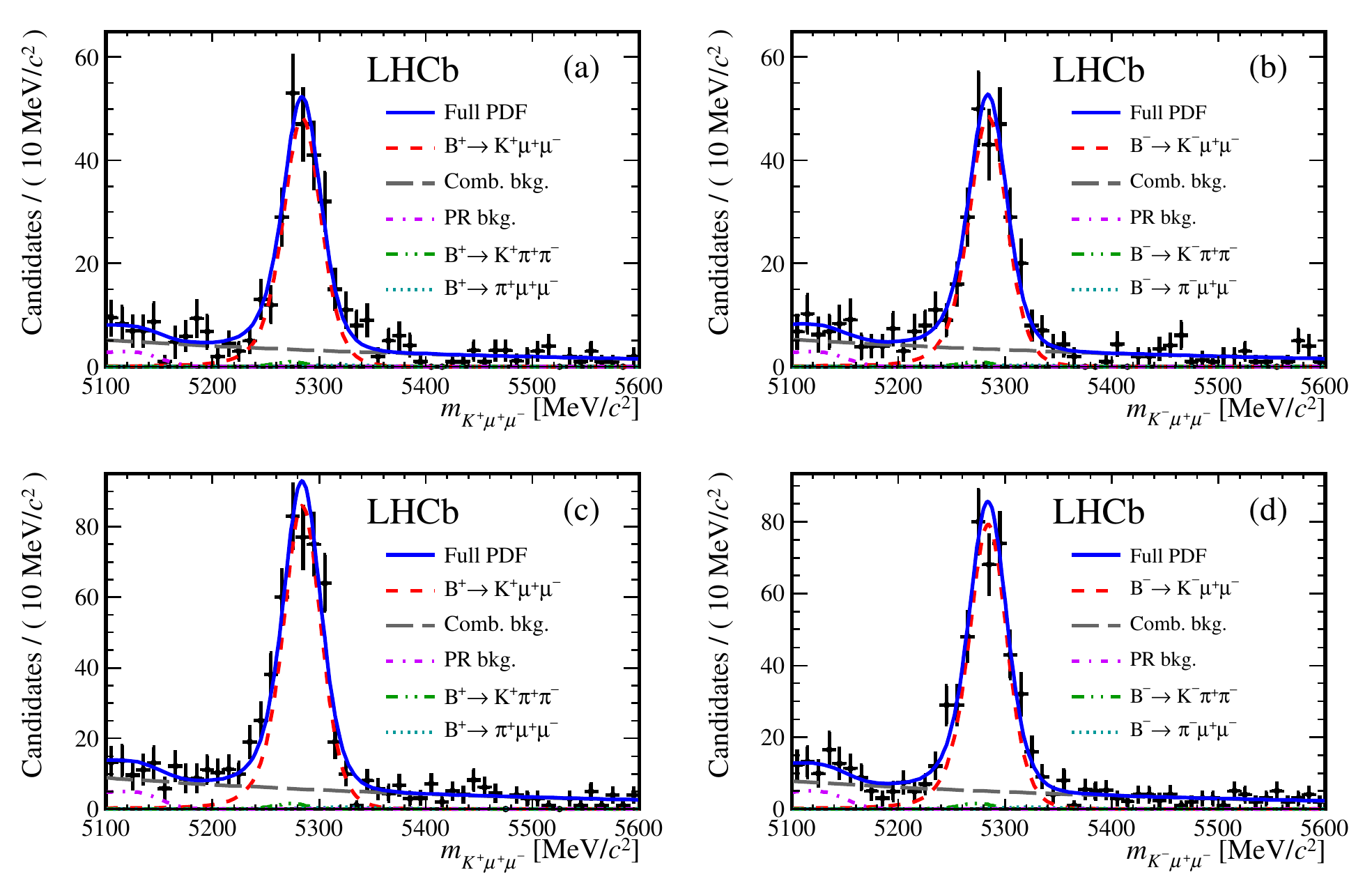}
\caption{\small Invariant mass distributions of \BToKmumu candidates for the full \qsq range. The results of the unbinned maximum likelihood fits are shown with blue, solid lines. Also shown are the signal component (red, short-dashed), the combinatorial background (grey, long-dashed), and the partially-reconstructed background (magenta, dot-dashed). The peaking backgrounds \BToKpipi (green, double-dot-dashed) and \BTopimumu (teal, dotted) are also shown under the signal peak. The four datasets are (a) \Bp and (b) \Bm for one magnet polarity, and (c) \Bp and (d) \Bm for the other.}

\label{fig:ACP4}
\end{figure}

In each \qsq bin, the \BToJpsiK and \BToKmumu data sets are divided according to the charge of the \Bp meson and magnet polarity, providing eight distinct subsets. These are fitted simultaneously with the parameters of the signal Cruijff function common for all eight subsets. For each subset, the only independent fitting parameters are the combined yield of the \Bp and \Bm decays and the values of \ARAW for the signal, control and background modes for each magnet polarity. The fits to the invariant mass distributions of the \BToKmumu candidates in the full \qsq range are shown in Fig.~\ref{fig:ACP4}.

The value of \ACP for each magnet polarity is determined from Eq.~\ref{eq:ACP2}, and an average with equal weights is taken to obtain a single value for the \qsq bin. To obtain the final value of \ACP for the full dataset, an average is taken of the values in each \qsq bin, weighted according to the signal efficiency and the number of \BToKmumu decays in the bin,

\begin{equation}
 \ACP = \frac{\sum^{7}_{i=1} (N_{i}\ACP^{i})/\epsilon_{i}}{\sum^{7}_{i=1}N_{i}/\epsilon_{i}},
\end{equation}
where $N_{i}$, $\epsilon_i$, and $\ACP^i$ are the signal yield, signal efficiency, and the fitted value of the \CP asymmetry in the $i\mathrm{th}$ \qsq bin.

Several assumptions are made about the backgrounds. The partially-reconstructed background is assumed to exhibit no \CP asymmetry. For \BTopimumu, \ACP is also assumed to be zero~\cite{PhysRevD.60.014005}. For the \BToKpipi decay, \ACP in each \qsq bin is taken from a recent \lhcb measurement~\cite{LHCb-PAPER-2013-027}. The effect of these assumptions on the result is investigated as a systematic uncertainty.

Various sources of systematic uncertainty are considered. The analysis relies on the assumption that the \BToKmumu and \BToJpsiK decays have the same final-state kinematic distributions, so that the relation in Eq.~\ref{eq:ACP2} is exact. To estimate the bias associated with this assumption, the kinematic distributions of \BToJpsiK decays are reweighted to match those of \BToKmumu, and the value of \ARAW is recalculated. The variables used are the momentum, \pt and pseudorapidity of the \Bp and \Kp mesons, as well as the \Bp decay time and the position of the kaon in the detector. The difference between the two values of \ARAW for each variable is taken as the systematic uncertainty. The total systematic uncertainty associated to the different kinematic behaviour of the two decays in each \qsq bin is calculated by adding each individual contribution in quadrature.

\begin{table}[tb]
\caption{Systematic uncertainties on \ACP from non-cancelling asymmetries arising from kinematic differences between \BToJpsiK and \BToKmumu decays, and fit uncertainties arising from the choice of signal shape, mass fit range and combinatorial background shape, and from the treatment of the asymmetries in the \BTopimumu and partially-reconstructed (PR) backgrounds. The total is the sum in quadrature of each component.}
  \begin{center}
\footnotesize
\begin{tabular}{c|D{.}{.}{2.4} D{.}{.}{2.4} D{.}{.}{2.4} D{.}{.}{2.4} D{.}{.}{2.4} D{.}{.}{2.4}|D{.}{.}{1.4}}
	&  \multicolumn{1}{c}{Residual} & \multicolumn{1}{c}{Signal} & \multicolumn{1}{c}{Mass} & \multicolumn{1}{c}{Comb.}  & \multicolumn{1}{c}{\ACP in} & \multicolumn{1}{c|}{\ACP in} & \\
\qsq bin (\gevgevcccc) & \multicolumn{1}{c}{asymmetries} & \multicolumn{1}{c}{shape} & \multicolumn{1}{c}{range} & \multicolumn{1}{c}{shape}  & \multicolumn{1}{c}{\BTopimumu}& \multicolumn{1}{c|}{PR}& \multicolumn{1}{c}{Total}\\
 \hline
$0.05<\qsq<2.00$\Tstrut& 0.005 & 0.005 & 0.002 & 0.002 & 0.004 & 0.002 & 0.008\\
$2.00<\qsq<4.30$  & 0.004 & 0.001 & 0.005 & 0.009  & 0.005 & 0.001 & 0.012\\
$4.30<\qsq<8.68$  & 0.001 & 0.001 & 0.001 & 0.001 & 0.005 & 0.002 & 0.005\\
$10.09<\qsq<12.86$ & 0.003 & 0.005 & 0.023 & 0.003  & 0.003 & 0.001 & 0.024\\
$14.18<\qsq<16.00$ & 0.006 & 0.001 & 0.004 & 0.003  & <0.001 & 0.001 & 0.008\\
$16.00<\qsq<18.00$ & 0.005 & 0.007 & 0.017 & <0.001  & <0.001& 0.001 & 0.019\\
$18.00<\qsq<22.00$ & 0.008 & 0.001 & 0.014 &  <0.001 & 0.001 & 0.001 & 0.016\\
\hline
 Weighted average\Tstrut& 0.001 & <0.001 & 0.003 & 0.001 & 0.003 & <0.001 & 0.005 \\
\hline
$1.00<\qsq<6.00$\Tstrut & 0.002 & <0.001 & 0.009 & 0.002 & 0.004 & 0.002 & 0.010 \\
  \end{tabular}\end{center}
\label{tab:AllSysts}
\end{table}

The choice of fit model also introduces systematic uncertainties. The fit is repeated using a different signal model, replacing the Cruijff function with the sum of two Crystal Ball functions~\cite{Skwarnicki:1986xj} that have the same mean and tail parameters, but different Gaussian widths. The difference in the value of \ACP using these two fits is assigned as the uncertainty. The fit is also repeated using a reduced mass range of ${5.17<m_{\NCKmumu}<5.60\gevcc}$ to investigate the effect of excluding the partially-reconstructed background. The difference in results obtained by modelling the combinatorial background using a second-order polynomial, rather than an exponential function, produces a small systematic uncertainty.

Uncertainties also arise from the assumptions made about the asymmetries in background events. Phenomena beyond the SM could cause the \CP asymmetry in \BTopimumu decays to be large~\cite{PhysRevD.60.014005}, and so the analysis is performed again for values of ${\ACP(\BTopimumu) = \pm0.5}$, with the larger of the two deviations in $\ACP(\BToKmumu)$ taken as the systematic uncertainty. As the partially-reconstructed background can arise from \BdToKstarmumu decays, the value of \ACP for this source background is taken to be $-0.072$~\cite{LHCb:2012kz}, the value from the \lhcb measurement, neglecting any further \CP violation in angular distributions. The difference in the fit result compared to the zero \ACP hypothesis is taken as the systematic uncertainty. Variations in $\ACP(\BToKpipi)$ have a negligible effect on the final result. A summary of the systematic uncertainties is shown in Table~\ref{tab:AllSysts}. The value of \ACP calculated by performing the fits on the data set integrated over \qsq is consistent with that from the weighted average of the \qsq bins.

\begin{figure}[htb]
 \centering
\includegraphics[width=0.7\textwidth]{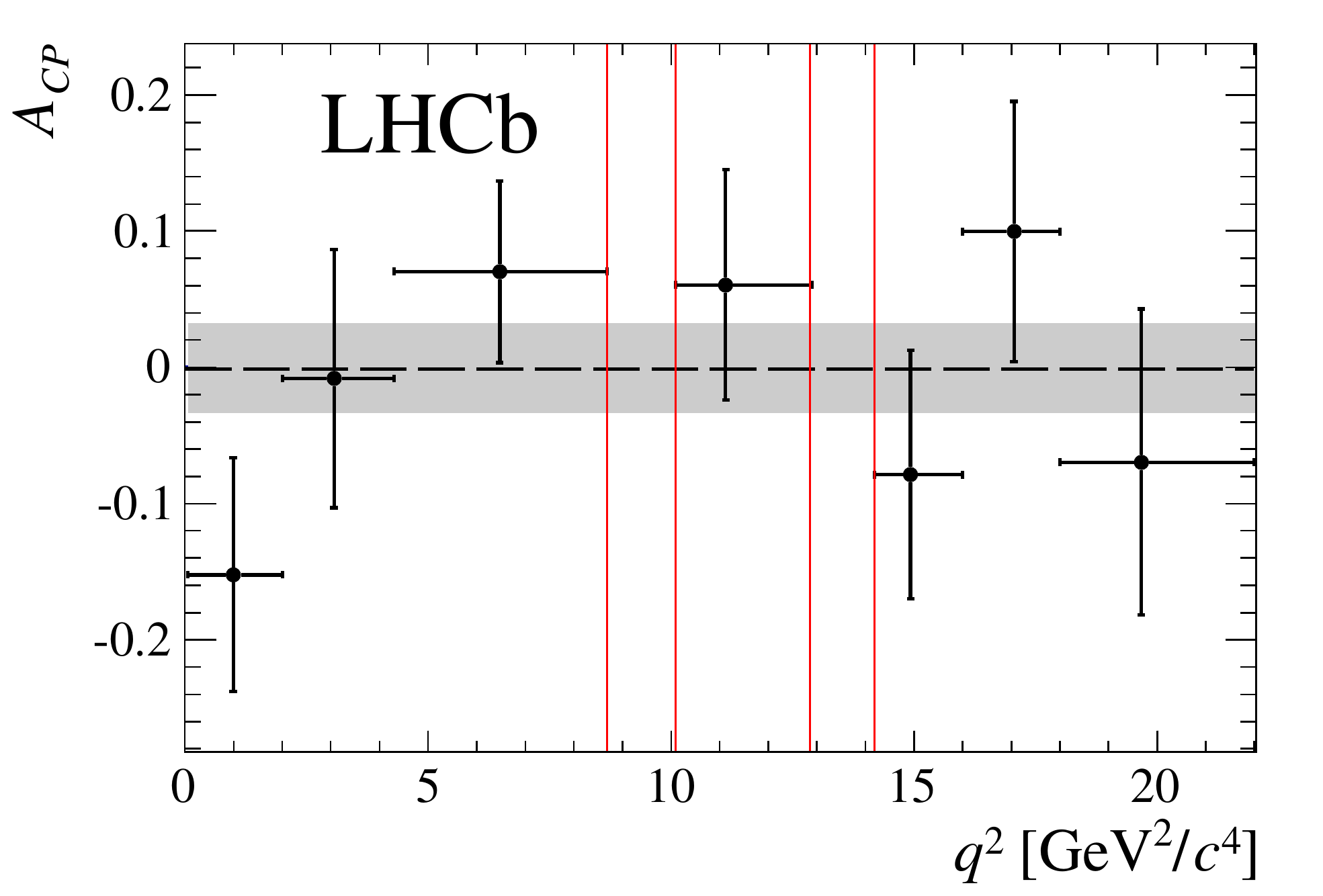}
\caption{Measured value of \ACP in \BToKmumu decays in bins of the \mumu invariant mass squared (\qsq). The points are displayed at the mean value of \qsq in each bin. The uncertainties on each \ACP value are the statistical and systematic uncertainties added in quadrature. The excluded charmonium regions are represented by the vertical red lines, the dashed line is the weighted average, and the grey band indicates the 1$\sigma$ uncertainty on the weighted average.}
\label{fig:ACP_Binned}
\end{figure}

The results for \ACP in each \qsq bin and the weighted average are displayed in Table~\ref{tab:results}, as well as in Fig.~\ref{fig:ACP_Binned}. The value of the raw asymmetry in \BToJpsiK determined from the fit is ${-0.016 \pm 0.002}$.
The \CP asymmetry in \BToKmumu decays is measured to be
\begin{equation*}
	\ACP = 0.000\pm 0.033\mbox{ (stat.)} \pm0.005 \mbox{ (syst.)} \pm 0.007\mbox{ }(\jpsi\Kp),
\end{equation*}
where the third uncertainty is due to the uncertainty on the known value of $\ACP(\BToJpsiK)$. This compares with the current world average of ${-0.05 \pm 0.13}$~\cite{PDG2012}, and previous measurements including the dielectron final-state~\cite{Babar:2012vwa,Wei:2009zv}. This result is consistent with the SM, as well as the \BdToKstarmumu decay mode, and improves the precision of the current world average for the dimuon mode by a factor of four. With the recent observation of resonant structure in the low-recoil region above the \psitwos resonance~\cite{LHCB-PAPER-2013-039}, care should be taken when interpreting the result in this region. Interesting effects due to physics beyond the SM are possible through interference with this resonant structure, and could be investigated in a future update of the measurement of \ACP.

\begin{table}[tb]
\centering
\footnotesize
\caption{Values of \ACP and the signal yields in the seven \qsq bins, the weighted average, and their associated uncertainties.}
\begin{tabular}{c|c D{.}{.}{1.4} D{.}{.}{1.4} D{.}{.}{1.4}}

 	                      &   &                                   & \multicolumn{1}{c}{Stat.}  & \multicolumn{1}{c}{Syst.}\\ 
  \qsq bin ($\gevgevcccc$) & \BToKmumu yield&\multicolumn{1}{c}{$\ACP \left( \BToKmumu \right)$} & \multicolumn{1}{c}{uncertainty} & \multicolumn{1}{c}{uncertainty}\\ \hline 
$0.05<\qsq<2.00$\Tstrut	& $164 \pm 14$  & -0.152 & 0.085 & 0.008  \\
$2.00<\qsq<4.30$	& $167 \pm 14$  & -0.008 & 0.094 & 0.012 \\
$4.30<\qsq<8.68$	& $339 \pm 21$  & 0.070 & 0.067 & 0.005 \\
$10.09<\qsq<12.86$	& $221 \pm 17$  & 0.060 & 0.081 & 0.024 \\
$14.18<\qsq<16.00$	& $145 \pm 13$  & -0.079 & 0.091 & 0.008 \\
$16.00<\qsq<18.00$	& $145 \pm 13$  & 0.100 & 0.093 & 0.019 \\ 
$18.00<\qsq<22.00$	& $120 \pm 13$  & -0.070 & 0.111  & 0.016\\ \hline
Weighted average\Tstrut	&  		& 0.000 & 0.033 & 0.005 \\  \hline
$1.00<\qsq<6.00$\Tstrut	& $362 \pm 21$	& -0.019 & 0.061 & 0.010 \\  \hline

\end{tabular}
\label{tab:results}
\end{table}

\section*{Acknowledgements}
We express our gratitude to our colleagues in the CERN
accelerator departments for the excellent performance of the LHC. We
thank the technical and administrative staff at the LHCb
institutes. We acknowledge support from CERN and from the national
agencies: CAPES, CNPq, FAPERJ and FINEP (Brazil); NSFC (China);
CNRS/IN2P3 and Region Auvergne (France); BMBF, DFG, HGF and MPG
(Germany); SFI (Ireland); INFN (Italy); FOM and NWO (The Netherlands);
SCSR (Poland); MEN/IFA (Romania); MinES, Rosatom, RFBR and NRC
``Kurchatov Institute'' (Russia); MinECo, XuntaGal and GENCAT (Spain);
SNSF and SER (Switzerland); NAS Ukraine (Ukraine); STFC (United
Kingdom); NSF (USA). We also acknowledge the support received from the
ERC under FP7. The Tier1 computing centres are supported by IN2P3
(France), KIT and BMBF (Germany), INFN (Italy), NWO and SURF (The
Netherlands), PIC (Spain), GridPP (United Kingdom). We are thankful
for the computing resources put at our disposal by Yandex LLC
(Russia), as well as to the communities behind the multiple open
source software packages that we depend on.

\ifx\mcitethebibliography\mciteundefinedmacro
\PackageError{LHCb.bst}{mciteplus.sty has not been loaded}
{This bibstyle requires the use of the mciteplus package.}\fi
\providecommand{\href}[2]{#2}

\end{document}